\documentclass[11pt,twoside]{article}
\usepackage{asp2014}

\aspSuppressVolSlug
\resetcounters

\begin{document}

\title{White Dwarf Mass Distribution}
\author{S. O. Kepler$^{1}$,
                D. Koester$^{2}$,
                A. D. Romero$^{1}$,
                G. Ourique$^{1}$, and
                I. Pelisoli$^{1}$
\affil{$^{1}$Instituto de F\'{\i}sica, Universidade Federal do Rio Grande do Sul, 91501-900  Porto-Alegre, RS, Brazil; \email{kepler@if.ufrgs.br}}
\affil{$^{2}$Institut f\"ur Theoretische Physik und Astrophysik, Universit\"at Kiel, 24098 Kiel, Germany; \email{koester@astrophysik.uni-kiel.de}}
}

\paperauthor{S.O. Kepler}{kepler@if.ufrgs.br}{0000-0002-7470-5703}{Universidade Federal do Rio Grande do Sul}{Instituto de F\'{\i}sica}{Porto Alegre}{RS}{91501-970}{Brazil}
\paperauthor{Detlev Koester}{koester@astrophysik.uni-kiel.de}{}{Universit\"at Kiel}{Institut f\"ur Theoretische Physik und Astrophysik}{Kiel}{Schleswig-Holstein}{24098}{Germany}
\paperauthor{Alejandra Daniela Romero}{aleromero82@gmail.com}{}{Universidade Federal do Rio Grande do Sul}{Instituto de F\'{\i}sica}{Porto Alegre}{RS}{91501-970}{Brazil}
\paperauthor{Gustavo Ourique}{ourique.gustavo@gmail.com}{}{Universidade Federal do Rio Grande do Sul}{Instituto de F\'{\i}sica}{Porto Alegre}{RS}{91501-970}{Brazil}
\paperauthor{Ingrid Pelisoli}{ingrid.pelisoli@gmail.com}{}{Universidade Federal do Rio Grande do Sul}{Instituto de F\'{\i}sica}{Porto Alegre}{RS}{91501-970}{Brazil}

\begin{abstract}
We present the mass distribution for all S/N$\geq$15 pure DA white dwarfs detected in the Sloan Digital Sky Survey  up to Data Release 12, 
fitted with Koester models for ML2/$\alpha=0.8$,  and with $T_\mathrm{eff} \geq 10\,000$~K, 
and for DBs with S/N$\geq$10, fitted with ML2/$\alpha=1.25$, for $T_\mathrm{eff}>16\,000$~K.  
These mass distributions are for $\log g\geq 6.5$ stars, i.e., excluding the Extremely Low Mass white dwarfs.
We also present the mass distributions corrected by volume with the $1/V_\mathrm{max}$ approach, for stars brighter than g=19. 
Both distributions have a maximum at $M=0.624~M_\odot$ but very distinct shapes.
From the estimated z-distances, we deduce a disk scale height of 300 pc. We also present 10 probable halo white dwarfs, from their galactic U, V, W velocities.

\end{abstract}
\section{Introduction}

Stars born with
initial masses up to 8.5--10.6~M$_{\sun}$ \citep{Woosley15}, corresponding to at least 95\% of all stars, become white dwarfs when they cannot fuse nuclear elements in the core anymore. For single star evolution, 
the minimum mass of the white dwarf is around 0.30--$0.45~M_\odot${} \citep[e.g.][]{Kilic07}. 
Considering the mass-radius relation of white dwarfs, this corresponds to a $\log g \geq  6.5$. 
Progenitors that would become lower mass white dwarfs live on the main sequence longer than the age of the Universe. 
We therefore determine our mass distribution only for white dwarfs with $\log g \geq 6.5$.

We estimated the masses of all DA white dwarfs found by \citet{Kleinman13}, \citet{Kepler15} and \citet{Kepler16a} among the 4.5 million spectra acquired by the Sloan Digital Sky Survey Data Release 12. For the mass
distribution we only consider spectra with S/N$\geq 15$ to have reliable mass determinations. 
The spectra were fitted with synthetic spectra from  model atmospheres of \citet{Koester10}, using ML2/$\alpha=0.8$ for DAs,
and  ML2/$\alpha=1.25$, for DBs.
We use the
mass--radius relations of \citet{Althaus05}, \citet{Renedo10} and \citet{Romero15},
to calculate the mass of our stars from the
$T_\mathrm{eff}$ and  $\log g$ values obtained from our fits,
after correcting to 3D convection following \citet{Tremblay13}.

\section{Mass Distribution}
Figure~\ref{dadb} shows the mass distribution by number for DAs with $T_\mathrm{eff}\geq 13\,000$~K, where convection is unimportant,
and for DBs with $T_\mathrm{eff}\geq 16\,000$~K reported by \citet{Koester15}.
Because our surface gravities show an unexplained decrease below $T_\mathrm{eff}=10\,000$~K, Figure~\ref{histv10m} shows the 
mass distribution for different cutoff temperatures.
\articlefigure[width=0.72\textwidth]{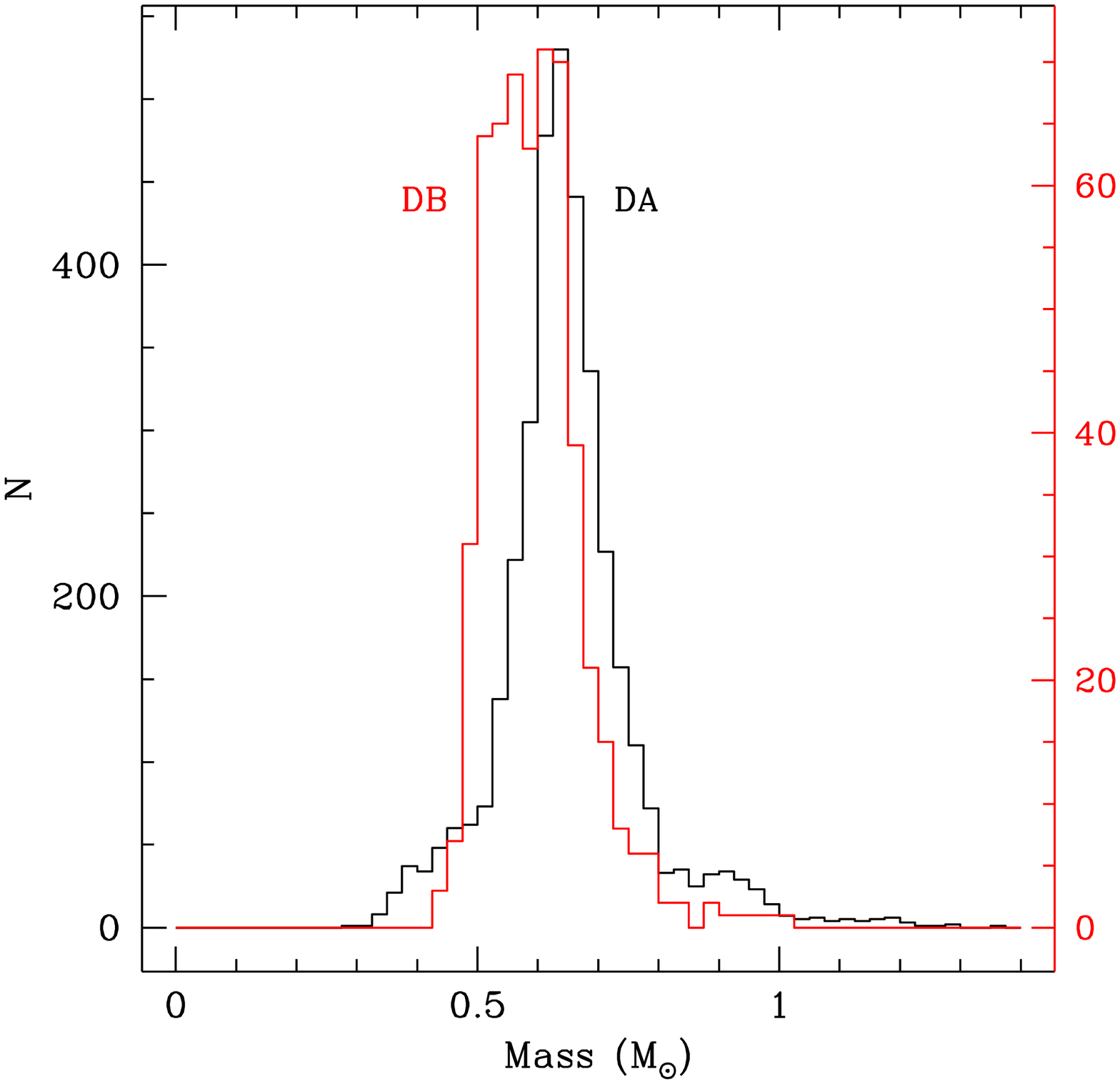}{dadb}{Mass distribution by number for 
3636 DAs with $T_\mathrm{eff}\geq 13000$~K, S/N$_g \geq 15$ and <S/N>=31 in black and
549 DBs with  $T_\mathrm{eff}\geq 16000$~K, S/N$_g \geq 10$ and <S/N>=21 in red.}

Considering white dwarfs with larger mass have smaller radius, and therefore can only be seen to smaller distances in a magnitude limited survey as SDSS, 
we calculated the density by correcting the visible volume with the $1/V_\mathrm{max}$ method of \citet{Schmidt68},
up to a maximum g=19 magnitude, shown in Figure~\ref{histv10m}.
The distribution shows that the DA and DB distributions have very different shapes. The DA's has a tail to larger masses, while the DB's is extended to lower masses. 
This is probably reflecting some limitation in the progenitors that can undergo very-late thermal pulses and become DBs.
\articlefigure[width=0.72\textwidth]{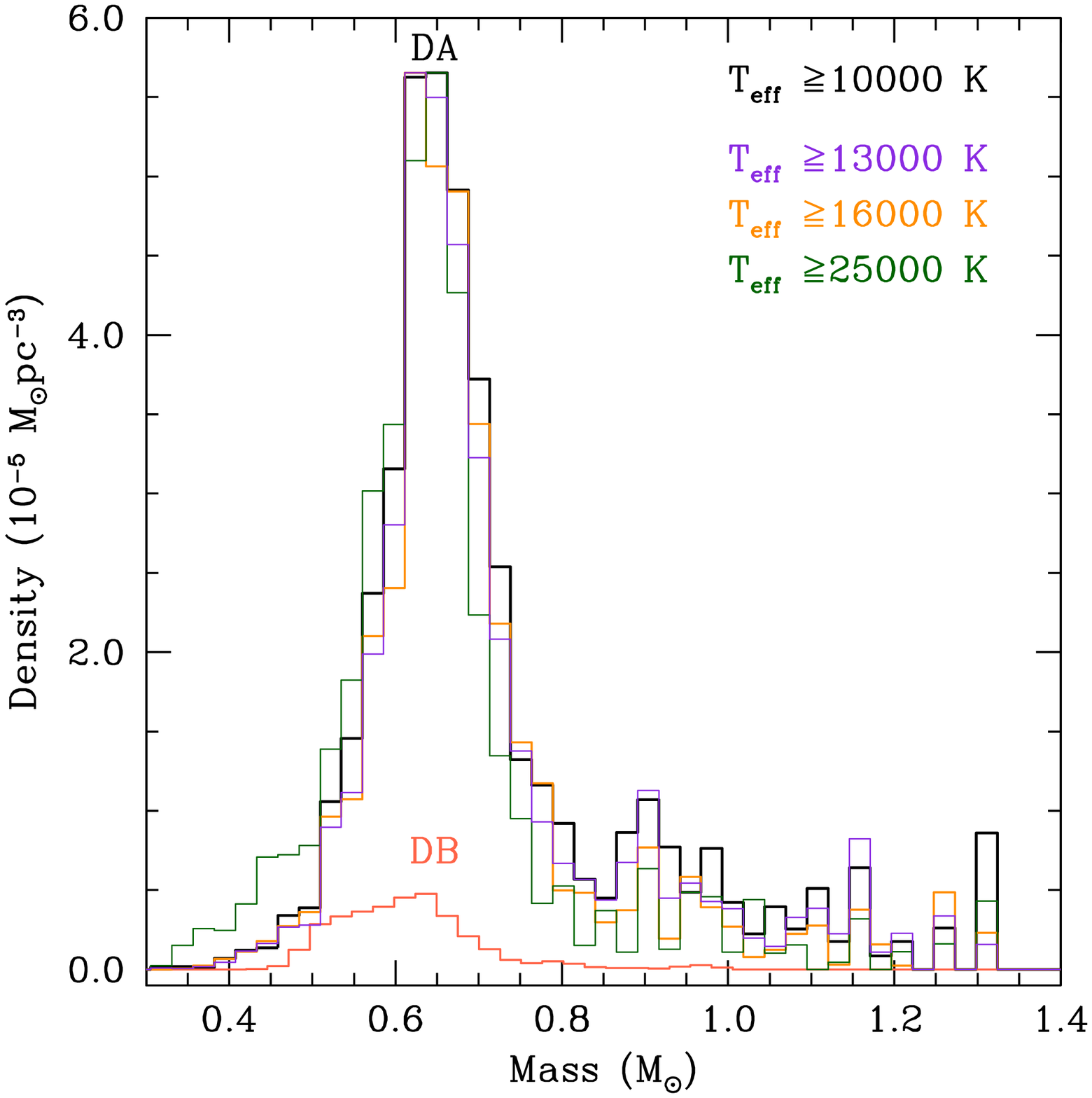}{histv10m}{Mass distribution corrected by the $1/V_\mathrm{max}$ method for DAs for different cutoff temperatures, and DB with $T_\mathrm{eff} \geq 16\,000$~K.
DAs with $T_\mathrm{eff} \geq 10000$~K, N=4054, <M>$=0.647\pm 0.002~M_\odot$ in black,
$T_\mathrm{eff} \geq 13000$~K, N=3637, <M>$=0.646\pm 0.002~M_\odot$ in violet,
$T_\mathrm{eff} \geq 16000$~K, N=3012, <M>$=0.641\pm 0.002~M_\odot$ in gold,
$T_\mathrm{eff} \geq 25000$~K, N=1121, <M>$=0.613\pm 0.003~M_\odot$ in green.}

\section{Discussion}
With our population synthesis analysis, we computed a theoretical mass distribution through a Monte Carlo simulation
fitting single star initial mass functions, initial-to-final mass relations for masses $0.45~M_\odot \leq  M  < 1.0~M_\odot$, to obtain a history of star formation
for the DAs with $T_\mathrm{eff} \geq 13\,000$~K.
Figure~\ref{histv13ps} shows the mean mass around 0.64~$M_\odot$ requires a burst of star formation in the last 2~Gyr,
as a white dwarf with such mass has a short lived progenitor mass with a mass around $2.5~M_\odot$.
This is in contrast with the uniform star formation estimated by \citet{Catalan08} from the ML2/$\alpha=0.6$ mass distribution of \citet{Kepler07}.
\articlefigure[width=0.72\textwidth]{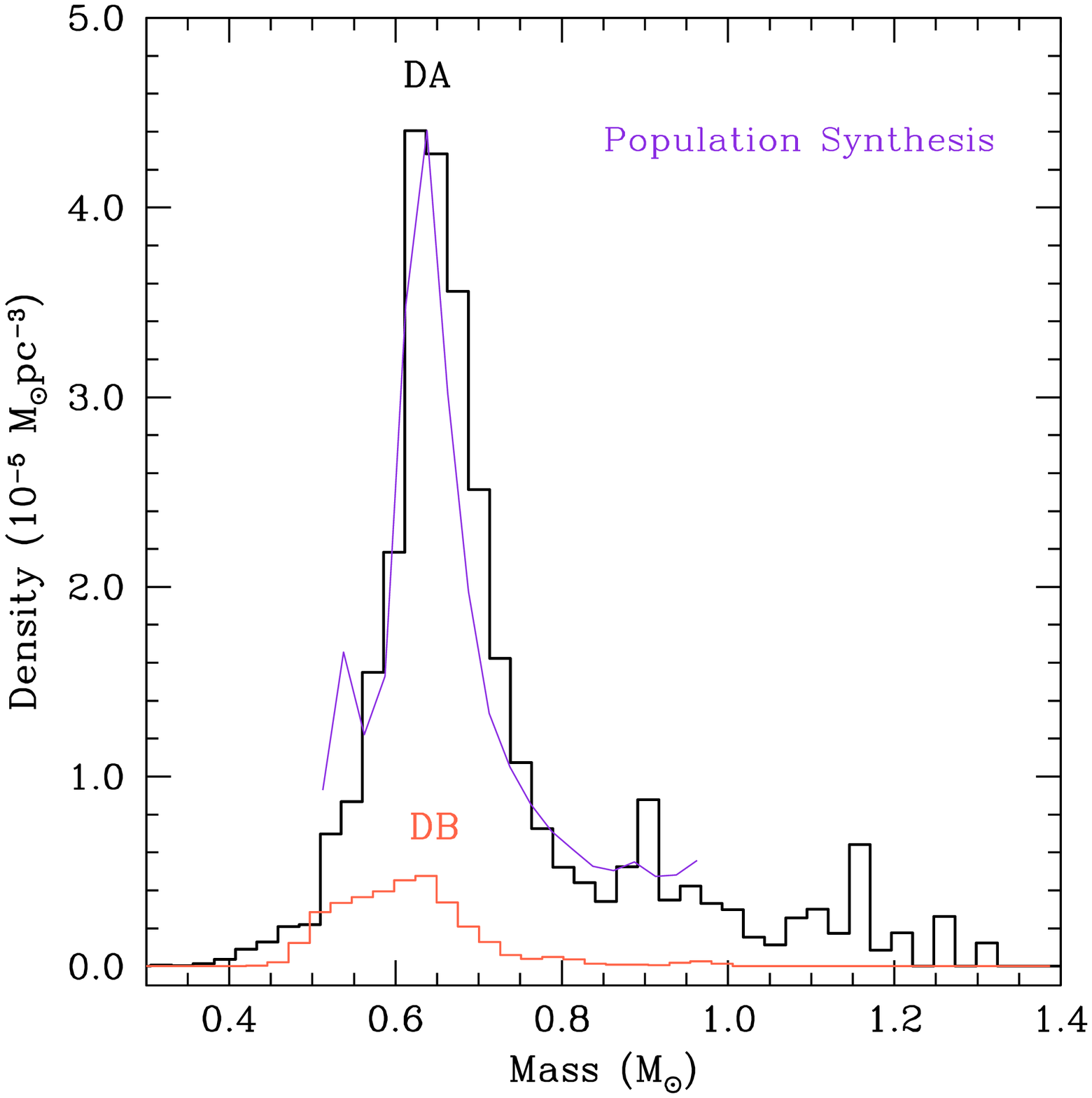}{histv13ps}{Mass distribution corrected by the $1/V_\mathrm{max}$ method for the
3637 DAs with $T_\mathrm{eff} \geq 13\,000$~K,
<M>$=0.646\pm 0.002~M_\odot$,
and DBs with $T_\mathrm{eff} \geq 16\,000$~K.
The blue line shows a population synthesis with a 30\% burst 2~Gyr ago, to account for the high mean mass.
The theoretical mass distribution represented by the population synthesis does not include either He-core or O-Ne-core models.}

Convolving the intensities from the theoretical models with filter transmission curves, and appropriate zero-points, we estimated the corresponding absolute
magnitudes. Comparing with the observed g-filter photometry we estimated the distance modulus,
obtaining the distances. From distances and the
galactic latitude, we estimated the distance of each star from the galactic plane z.
Figure~\ref{z} shows the distance above the galactic plane for each star studied, showing the disc scale height is around 300~pc
for DAs and a few parsecs larger for DBs.
\articlefigure[width=0.72\textwidth]{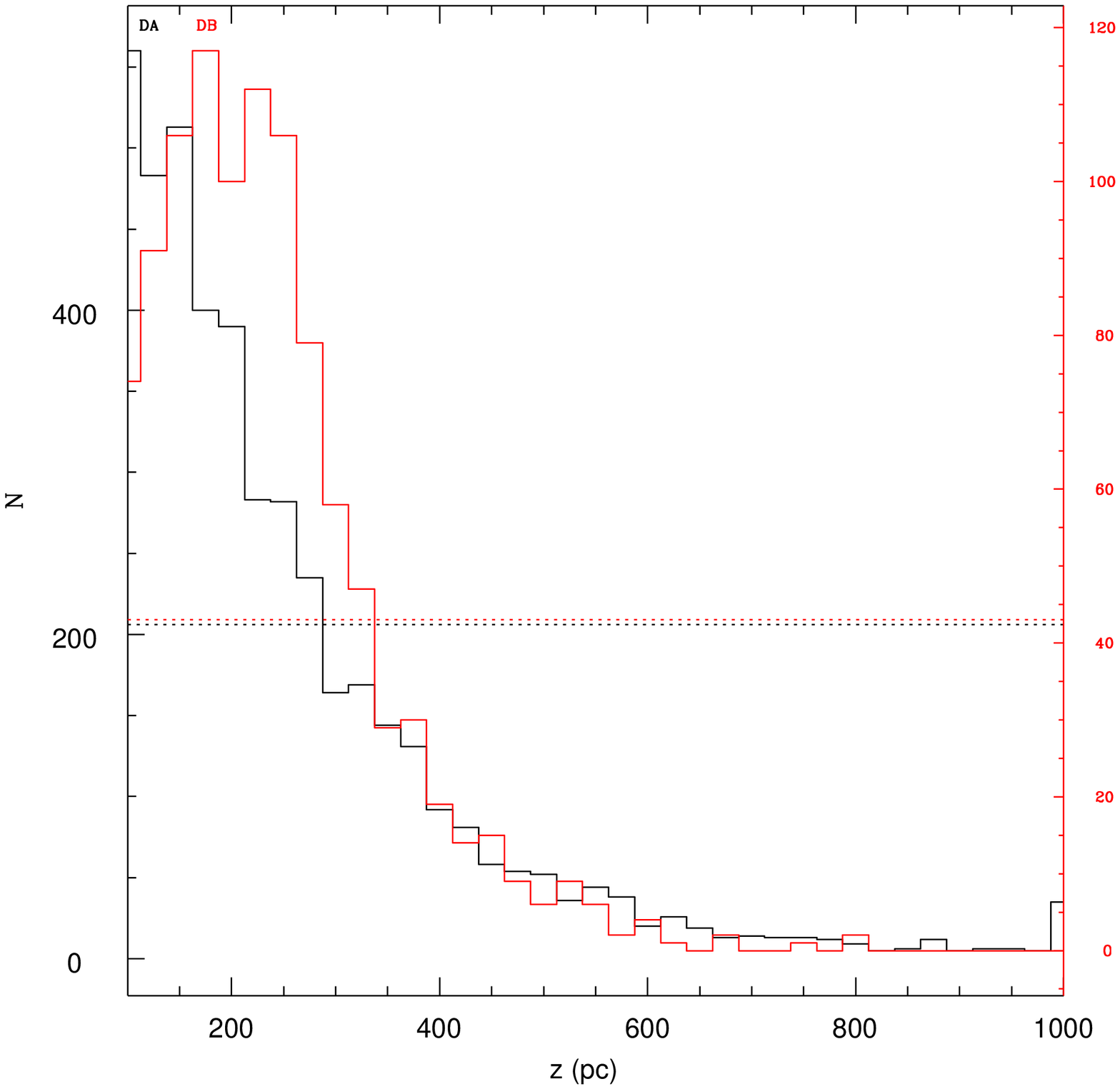}{z}{Histogram of the distribution of DAs and DBs versus the z distance above the galactic plane. The 1/e line drawn shows the scale height for the plane is around 300~pc
for DAs and a few parsecs larger for DBs.}

Finally, using the distances, our measured radial velocities and the proper motions obtained from APOP \citep{APOP} only for those stars which has a measured proper motion larger than three times its uncertainty, we estimated the galactic velocities U, V, and W for each star \citep[e.g][]{Johnson87}.
We compared the proper motions from APOP \citep{APOP} with those of \citet{Munn14} and they are very similar. 
In Figure~\ref{uvws}, we show the galactic velocities we infer for each star.
As expected, most white dwarfs observed by SDSS belong to the thin and thick disk. Because of the saturation limit around $g=14.5$, 
nearby white dwarfs only if
very cool 
are included.
The SDSS observations are also preferentially for directions across the galactic disk.

In Table~\ref{halo} we list the 10 stars with galactic velocities outside the thin and thick disk ellipsis of \citet{Kordopatis11}, which are probably halo white dwarfs,
or the result of a binary interaction.
\articlefigure[width=0.72\textwidth]{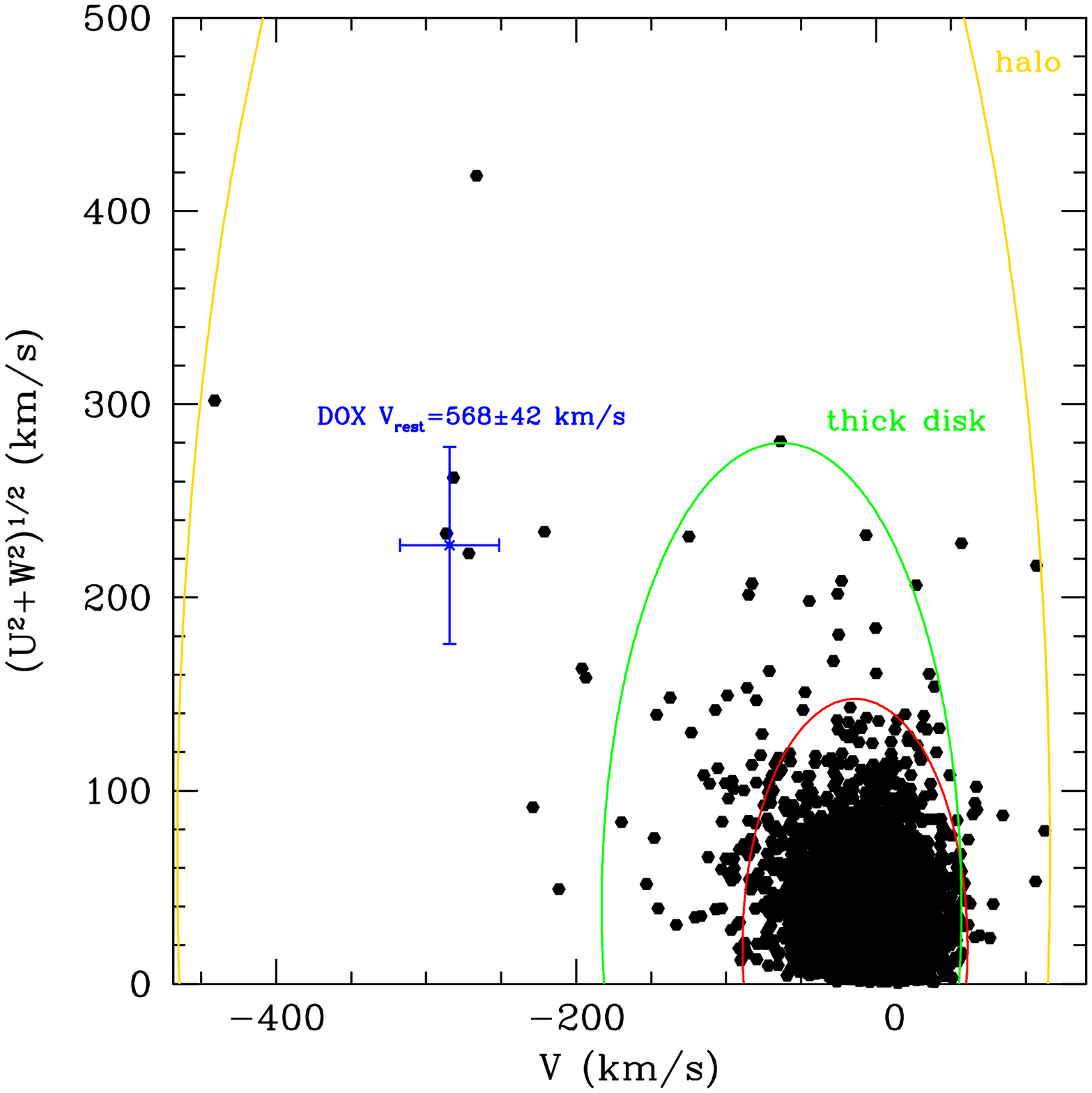}{uvws}{Galactic velocities obtained from the radial velocity, proper motion and distance modulus for each DA white dwarf. The ellipsis plotted are the $3\sigma$ mean velocities of
stars in the thin disk, thick disk and halo \citep{Kordopatis11}.
The blue cross labeled DOX represents the oxygen atmosphere white dwarf found by \citet{Kepler16b} and its uncertainties can be used as reference.}
\begin{table}[!ht]
\caption{DA white dwarfs for which their galactic velocities indicate probable halo members.\label{halo}}
\smallskip
\begin{center}
\begin{tabular}{lccccrccrr}
\tableline
SDSS J              &S/N  & g    &$\sigma_g$&$T_\mathrm{eff}$&$\sigma_T$&$\log g$&$\sigma_{\log g}$&dist&z\ \ \cr
                     &&   (mag)   &(mag)     &(K)             &(K)       &(cgs)   &(cgs)             &(pc)&(pc)\cr
\noalign{\smallskip}
\tableline
\noalign{\smallskip}
081514.42+511311.39  &  33 & 17.647 & 0.014 & 74528 &   644 &  7.354 & 0.027 & 1142 &   634\cr
091734.49+020924.37  &  15 & 18.882 & 0.015 & 16261 &   268 &  7.660 & 0.053 &  448 &   24\cr
113219.73-075441.94  &  22 & 19.068 & 0.033 & 33820 &   256 &  7.240 & 0.057 & 1431 &  109\cr
115045.04+191854.61  &  15 & 19.112 & 0.026 & 19135 &   247 &  8.122 & 0.040 &  407 &   39\cr
121731.31+610520.36  &  31 & 18.075 & 0.017 & 41495 &   443 &  7.895 & 0.038 &  655 &   54\cr
123827.80+312138.30  &  26 & 17.743 & 0.022 & 60285 &  1551 &  7.730 & 0.080 &  802 &   79\cr
125816.99+000710.24  &  20 & 18.135 & 0.015 & 14910 &   159 &  7.870 & 0.035 &  256 &   22\cr
152658.83+021510.19  &  17 & 18.871 & 0.016 & 48853 &  1297 &  7.200 & 0.098 & 1739 &  123\cr
225513.66+230944.14  &  36 & 17.771 & 0.024 & 30007 &   118 &  7.522 & 0.020 &  529 &   28\cr
230228.08+231747.90  &  21 & 19.441 & 0.010 & 40716 &   720 &  7.740 & 0.064 & 1054 &   57\cr
\noalign{\smallskip}
\tableline\
\end{tabular}
\end{center}
\end{table}
\acknowledgements SOK, ADR, GO and IP are supported by CNPq-Brazil. DK received support from program Science without Borders, MCIT/MEC-Brazil. This research has made use of NASA's Astrophysics Data System and of the cross-match service provided by CDS, Strasbourg. Funding for the Sloan Digital Sky Survey has been provided by the Alfred P. Sloan Foundation, the U.S. Department of Energy Office of Science, and the Participating Institutions. The SDSS web site is www.sdss.org.

\end{document}